\documentclass[12pt,preprint,a4paper]{aastex}
\usepackage{graphics,epsf}
\usepackage{amsmath}                
\usepackage{amsfonts}               
\usepackage{amssymb}                
\usepackage{epsfig}                 

\def \s{~\rm{s}}
\def \km{~\rm{km}}

\def \yr{~\rm{yr}}

\def \days{~\rm{day}}

\def \mum{~\rm{\mu m}}

\def \astrobj#1{#1}

\begin{document}

\title{PREDICTION FOR THE HE I $\lambda 10830 \rm{\AA}$ ABSORPTION
WING IN THE COMING EVENT OF ETA CARINAE}

\author{Amit Kashi\altaffilmark{1} and Noam Soker\altaffilmark{1}}

\altaffiltext{1}{Department of Physics, Technion$-$Israel Institute of
Technology, Haifa 32000 Israel; kashia@physics.technion.ac.il;
soker@physics.technion.ac.il.}



\begin{abstract}
We propose an explanation to the puzzling appearance of a wide blue absorption wing
in the He~I~$\lambda10830 \rm{\AA}$ P Cygni profile of the massive binary star $\eta$ Car
several months before periastron passage.
Our basic assumption is that the colliding winds region is responsible for the
blue wing absorption.
By fitting observations, we find that the maximum outflow velocity of this absorbing
material is $\sim 2300 \km \s^{-1}$.
We also assume that the secondary star is toward the observer at periastron passage.
With a toy-model we achieve two significant results.
(1) We show that the semimajor axis orientation we use can account for
the appearance and evolution of the wide blue wing under our basic assumption.
(2) We predict that the Doppler shift (the edge of the absorption profile)
will reach a maximum $0-3~{\rm weeks}$ before periastron passage, and
not necessarily exactly at periastron passage or after periastron passage.
\end{abstract}


\section{INTRODUCTION}
\label{sec:intro}

There is no boring feature when it comes to the massive binary system
\astrobj{$\eta$ Car}.
Its $P=5.54 \yr$ ($P =2022.7 \pm 1.3~$d; Damineli et al. 2008a) periodicity has been
observed in the radio (Duncan \& White 2003), IR (Whitelock et al. 2004),
visible (e.g., van Genderen et al. 2006), X-ray (Corcoran 2005),
and in many emission and absorption lines (e.g., Damineli et al. 2008a,b).
Every orbital cycle the system experiences a spectroscopic event,
defined by the fading, or even disappearance, of high-ionization
emission lines (e.g., Damineli 1996; Damineli et al. 1998, 2000,
2008a,b; Zanella et al. 1984). The rapid changes in the continuum,
lines, and in the X-ray properties (e.g., Martin et al. 2006,a,b;
Davidson et al. 2005; Nielsen et al. 2007; van Genderen et al. 2006;
Damineli et al. 2008b; Corcoran 2005) are assumed to occur near the
periastron passages of the highly eccentric, $e \simeq 0.9$, binary
orbit (e.g., Hillier et al. 2006).

The He~I~$\lambda10830 \rm{\AA}$ high excitation line has a complex P Cygni profile,
composed of three blue-shifted peaks with significant variations over the cycle
(Damineli et al. 1998; 2008b).
The emission profile has significant variations over the cycle.
The Doppler shifts of the peaks are of relatively low velocities,
$\vert v_{\rm peaks} \vert < 300 \km \s^{-1}$ (Damineli et al. 2008b).
The location of the minimum of the profile (the deepest point in absorption) does not
change with orbital phase, and stays at $v_{obs-m}=-570 \km \s^{-1}$.
However, the absorption profile does change.
In particular, just before periastron a wide blue wing appear in absorption, reaching
$\sim -1000 \km \s^{-1}$ a month before periastron, and $\sim -1800 \km \s^{-1}$ at periastron,
and the maximum equivalent width of absorption occurs $10 \days$ after periastron
passage (Damineli et al. 2008b).
Damineli et al. (2008b) approximated the average radial velocity of the
absorption profile at half intensity, and
found it to change from $-640 \km \s^{-1}$ before phase zero
to $-450 \km \s^{-1}$ shortly after phase zero.

An additional He~I~$\lambda10830 \rm{\AA}$ blue absorption feature of up to
$-1000 \km \s^{-1}$, is observed at several arcseconds from the center in the lobes (Smith 2002).
In this paper we refer only to the He~I~$\lambda10830 \rm{\AA}$ ground measurements
of Damineli et al. (2008b).
We note that other absorption and emission lines of He~I can be formed
in different regions in the binary system (see also Kashi \& Soker 2007b).
In particular, some visible He~I lines can be formed in the hot winds of the two
stars close to their origin,
as compared with the He~I~$\lambda10830 \rm{\AA}$ that is formed in cooler regions.
We show that this cooler region can be the post-shock primary wind.
Therefore, different He~I lines need not have the same behavior along the orbit.

Because of the winds' very complicated flow structure, when starting this project we limited
ourself to build a toy-model in order to achieve two goals:
(1) To show that the orientation where the secondary is toward us at
periastron ($\omega = 90^\circ$), can be accounted for
the development of a wide blue absorption wing starting several weeks
before periastron passage.
(2) To encourage a nightly observation of the
He~I~$\lambda10830 \rm{\AA}$ close to periastron passage.

\section{THE STARS AND THEIR WINDS}
\label{sec:stars}

The $\eta$ Car binary parameters used by us are compiled by using results from
several different papers (e.g., Ishibashi et al. 1999; Damineli et al. 2000; Corcoran et
al. 2001, 2004; Hillier et al. 2001; Pittard \& Corcoran 2002;
Smith et al. 2004; Verner et al. 2005).
The assumed stellar masses are $M_1=120 M_\odot$, $M_2=30 M_\odot$,
the eccentricity is $e=0.9$, and the orbital period is $P=2024 \days$.
The mass loss rates and terminal speeds are $\dot M_1=3 \times 10^{-4} M_\odot
\yr^{-1}$, $\dot M_2 =10^{-5} M_\odot \yr^{-1}$, $v_{\rm 1,\infty}=500 \km \s^{-1}$ and
$v_{\rm 2,\infty}=3000 \km \s^{-1}$.
For these parameters, the half opening angle of the wind-collision
cone is $\phi_a \simeq 60 ^\circ$ (Akashi et al. 2006); because of the orbital
motion this angle is not constant, and we calculate it along the orbit in the present paper.
For the inclination angle we take $i = 41^\circ$ (Davidson et al. 2001; Smith 2002, 2006).

\begin{figure}[!t]
\resizebox{0.89\textwidth}{!}{\includegraphics{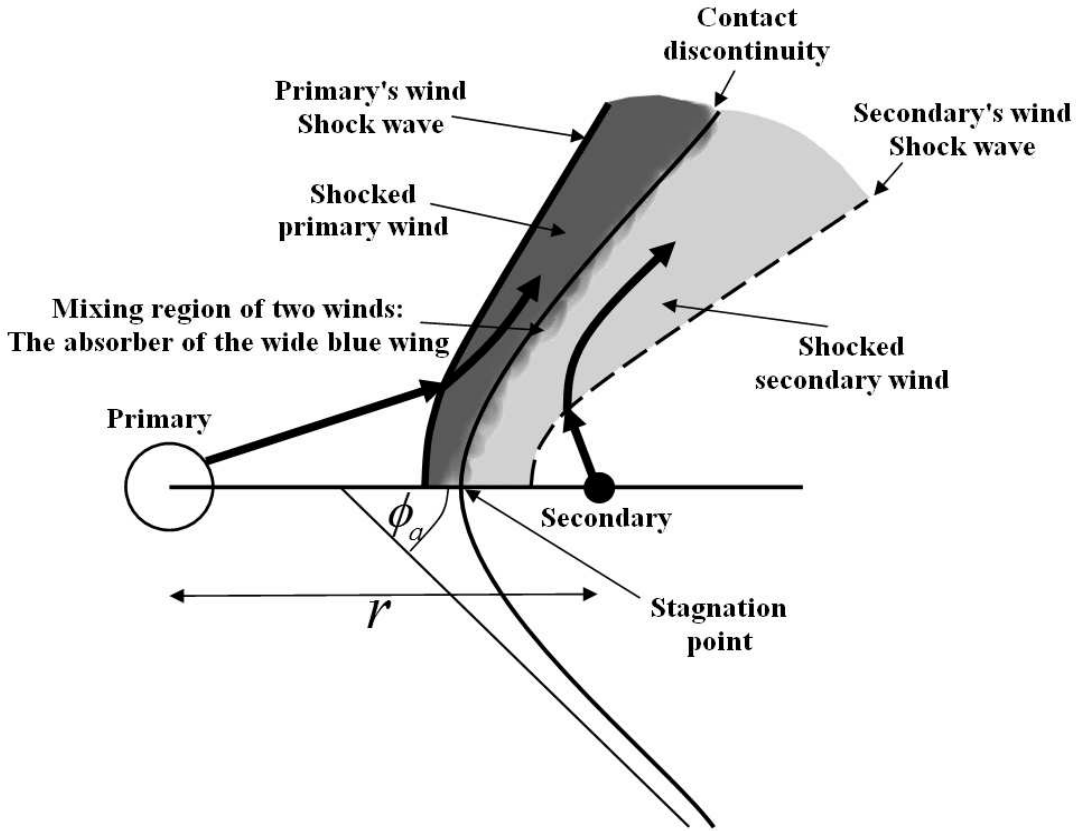}}
\caption{\footnotesize
The colliding winds structure. The primary's and secondary's winds
mix close to the contact discontinuity.
According to our models this region is the absorber of the wide
He~I~$\lambda10830 \rm{\AA}$ blue wing.
The  half opening angle of the wind-collision cone is
$\phi_a \simeq 60 ^\circ$.
The thick arrows indicate the flow lines.}
\label{fig:colliding}
\end{figure}

The primary's wind speed depends on latitude (Smith et al. 2003).
The minimum in the He~I~$\lambda10830 \rm{\AA}$ line profile suggests that
the primary's wind speed toward us is $v_{obs-m}=-570 \km \s^{-1}$.
The two winds collide, and form a flow structure, schematically drawn in
Figure \ref{fig:colliding}. The two winds go through two respective shock waves,
and form a contact discontinuity between them. The contact discontinuity asymptotically forms
a conical shell surface. The radiative cooling time of the post-shocked primary's wind is short.
The post-shocked primary's wind forms a dense flow along the contact discontinuity, which we
refer to as the \emph{conical shell.}
Absorption is expected to take place mainly outside the stagnation-point region,
and so we approximate the conical shell as an ideal cone.

The He~I~$\lambda10830 \rm{\AA}$ absorption profile results from He~I
atoms which absorb from the He~I~$\lambda10830 \rm{\AA}$ emission line and
from the continuum emitted by dust. As we are more interested in the
wide wing, the continuum is more relevant to our study.
As the system approaches periastron, dust is formed closer and closer
to the binary system (Kashi \& Soker 2008a).
This makes things very complicated, as the conical shell,
where the absorbing gas reside according to our model,
also changes its size. Some of the dust is formed in the conical shell itself.

Other lines, e.g. $H\alpha$ (Smith et al. 2003),
also show some fast blueshifted absorption wings.
Some of these wings disappear when the system approaches periastron.
This is not a problem for our model,
since those lines originate in different regions than where
the He~I~$\lambda10830 \rm{\AA}$ line does
(e.g. the $H\alpha$ comes from the primary's wind; Smith et al. 2003).
We emphasize again that some He~I lines (in particular in the visible
band) are formed in different regions than the He~I~$\lambda10830 \rm{\AA}$ line,
and their behavior is different as well.

The binary parameter that is most controversial is the orientation of the semimajor
axis$-$the periastron longitude.
Some researchers argue that the secondary (less massive)
star is away from us during periastron passages, i.e. an orbital longitude
of $\omega = 270^\circ$ (e.g., Nielsen et al. 2007; Damineli et al. 2008b),
others argue that the secondary is toward us during periastron passages,
$\omega = 90^\circ$ (Falceta-Gon\c{c}alves et al. 2005; Abraham et al.
2005; Kashi \& Soker 2007b, 2008b), and still different values exist in the literature
(Davidson 1997; Smith et al. 2004; Dorland 2007; Henley et al. 2008; Okazaki et al. 2008a).
Following our recent paper (Kashi \& Soker 2008b) we will take the orientation to
be such that the secondary is toward us at periastron ($\omega = 90^\circ$).

\section{THE TOY-MODEL}
\label{sec:toy}

We suggest that the conical shell is responsible for the absorption
of the He~I~$\lambda10830 \rm{\AA}$ high excitation line.
When the two winds meet at the contact discontinuity,
part of the shocked fast secondary's wind is mixed
with part of the shocked slow wind (Pittard 2007),
and accelerates part of it to higher velocities.
The conical shell has an asymptotic angle of $\phi_a\simeq60^\circ$,
which we take to be the flow direction of the absorbing gas.

In order to make the model more accurate we will take into account the time
dependance of some parameters.
The primary's wind velocity profile can be described using the $\beta$-profile:
\begin{equation}
v_1(r)=v_s+(v_{\rm 1,\infty}-v_s)\left(1-\frac{R_1}{r}\right)^{\beta} ,
\label{eq:v1}
\end{equation}
where $v_s=20 \km \s^{-1}$ is the sound velocity on the primary's
surface, $v_{\rm 1,\infty}=500 \km \s^{-1}$ is the primary's wind
terminal velocity, and $\beta=1$ is a parameter of the wind model.

The radial (along the line joining the two stars) component of the
relative velocity between the secondary star and the primary's wind is
$v_1-v_r$, where $v_r$ the radial component of the orbital velocity;
$v_r$ is negative when the two stars approach each other. The total
relative speed between the secondary and the primary's wind is
\begin{equation}
v_{\rm wind1} = \left[v_\theta^2 + (v_1-v_r)^2 \right]^{1/2},
\label{eq:vwind1}
\end{equation}
where $v_\theta$ is the tangential component of the orbital velocity.
The orbital motion and the variation of the primary wind speed with distance
from the primary have a small influence on the conical shell asymptotic angle $\phi_a$.
We will use the expression given by Eichler \& Usov (1993)
\begin{equation}
\phi_a \sim 2.1
\left(1-\frac{\eta^{\frac{4}{5}}}{4}\right)\eta^{\frac{2}{3}} ,
\label{eq:phia}
\end{equation}
where
\begin{equation}
\eta \equiv \sqrt{\frac{\dot M_2 v_{\rm 2,\infty}}{\dot M_1 v_{\rm wind1}}} .
\label{eq:eta}
\end{equation}

\begin{figure}[!t]
\resizebox{0.89\textwidth}{!}{\includegraphics{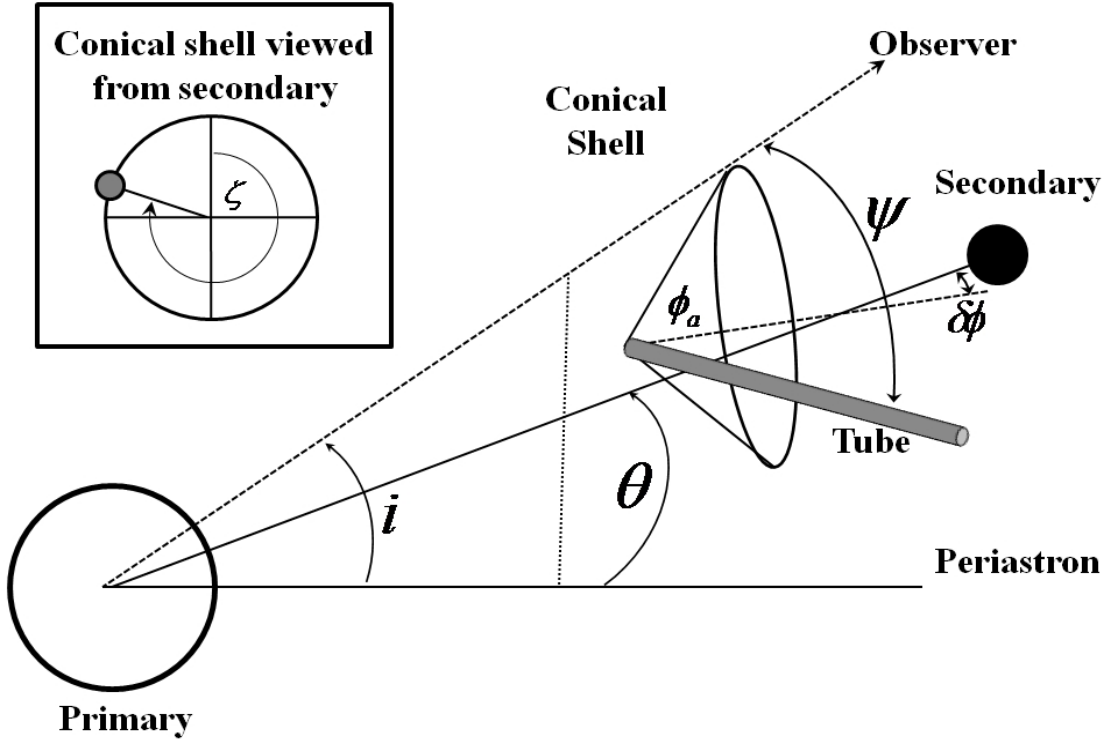}}
\caption{\footnotesize
The geometry of the toy-model and definition of several parameters.
The geometry is plotted near periastron ($\theta=0$),
where according to our model the secondary is toward us.
An example of a ``tube'' is also shown.
Both $\theta$ and $\delta \phi$ are measured in the equatorial plane.
At each point along the orbit there is one value for each
of the variables $\theta$, $\delta \phi$, and $\phi_a$, while we integrates over many tubes on
the conical shell, each with its value of direction to the observed $\psi$.}
\label{fig:cone}
\end{figure}

\begin{figure}[!t]
\resizebox{0.89\textwidth}{!}{\includegraphics{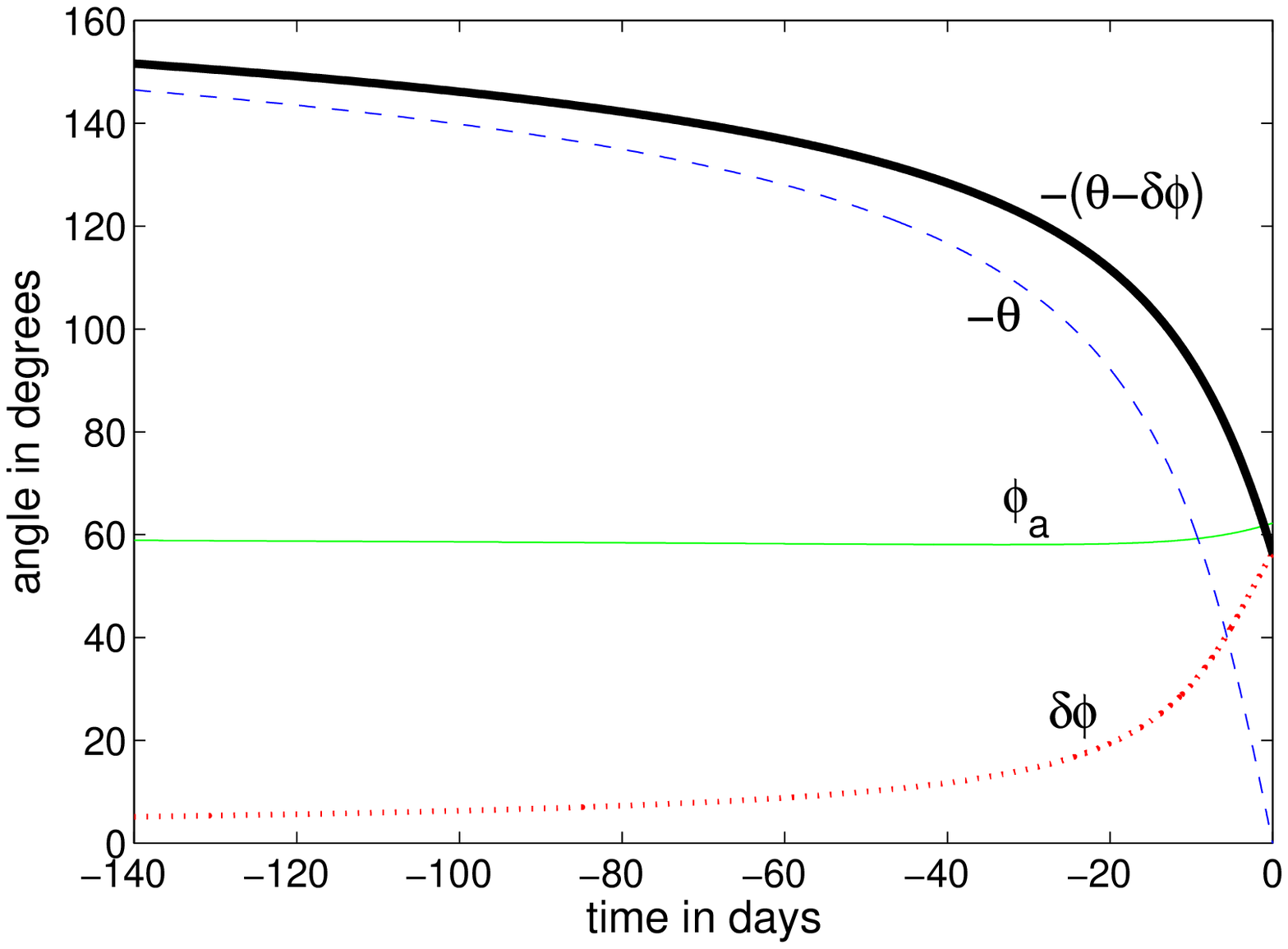}}
\caption{\footnotesize
The time variation of angles defined in the text.
$\theta$ is the orbital angle (line from primary to secondary) with $\theta=0$ at periastron.
$\delta \phi$ is the angle between two lines from the secondary: one to the primary star,
and one to the stagnation point of the colliding winds.
Both $\theta$ and $\delta \phi$ are measured in the equatorial plane.
$\phi_a$ is the opening angle of the cone, defined around the
cone axis. }
\label{fig:angles}
\end{figure}

We will take into consideration the rotation of the cone relatively to the line
connecting the two stars.
This rotation occurs due to the orbital velocity of the conical shell,
and has a considerable influence close to periastron.
We define $\delta \phi$ to be the angle measured from the secondary between the
direction to the primary and that to the stagnation point (see Soker 2005 for further details)
\begin{equation}
\cos (\delta \phi)=\frac{v_1-v_r}{v_{\rm wind1}}
\label{eq:deltaphi}
\end{equation}
We find that close to periastron $\delta \phi \simeq 56^\circ$.
The geometry and different parameters are shown in Figures \ref{fig:colliding}, \ref{fig:cone} and \ref{fig:angles}.


A closely-related geometry was used by Hill et al. (2000) to fit the excess emission
observed in the C III $\lambda 5696 \rm{\AA}$ line in the spectra of WR 42 and WR 79.
Luehrs (1997) also used a somewhat different geometry to fit the excess emission observed in the
C III $\lambda 5696 \rm{\AA}$ line in the spectra of WR 79.
These two models were purely geometric.
In the model we suggest for the absorption, some more considerations have to be taken in to account.

Although the absorbing gas is a continuous media, we decompose the conical shell into `tubes'.
We define $\zeta$ to be the azimuthal angle on the surface of the conical shell;
$\zeta=0$ in a direction perpendicular to the equatorial plane, and is measured clockwise.
Tubes exist from $\zeta=0$ to $\zeta=2 \pi$.
For each tube on the conical shell we calculate the angle $\psi$ between the tube
and the observer as a function of the orbital angle $\theta$ (see Figure \ref{fig:cone})
\begin{equation}
\cos\psi=\cos\phi_a\sin i \cos(\theta - \delta \phi) + \sin \phi_a \sin \zeta \sin i \sin(\theta - \delta \phi)
+ \sin\phi_a \cos \zeta \cos i .
\label{eq:psi}
\end{equation}

We take $v_0 < v_{\rm m} <v_{\rm max}$ to be the maximum velocity attended by a mass
element, $dm$, in a tube; in each tube there is a large number of mass elements formed by
the mixing process of the two winds.
By mixing we refer also to secondary shocked wind segments that have been cooled
to temperature low enough to contribute to the absorption at high speeds.
They reside near the contact discontinuity as well.
Here $v_0 = v_{\rm 1,\infty} = 500 \km \s^{-1}$ is the primary's wind velocity, and
$v_{\rm max}$ is a parameter of the model that is constraint to be
$v_{\rm max} \la v_{\rm 2,\infty}=3000 \km \s^{-1}$.
The projected velocity of mass element having a velocity $v_{\rm m}$ and residing
in a tube with an angle of $\psi$ to our line of sight is $v_D=v_{\rm m} \cos\psi$.

Consider one of our tubes; it has a length $L$ and a circular cross section with a radius $R_t$,
such that $R_t \ll L$. The effective cross section of the absorbing tube is $2R_t L \sin \psi$.
Therefore, the contribution of each tube to absorption is multiplied by $\sin \psi$.
The amount of accelerated primary's wind to each velocity $v_{\rm m}$ is hard to predict.
A numerical simulation beyond the scope of this work needs to be done in order to determine this amount.
We simply take the velocity distribution to be constant, namely, in each tube the
fraction of the gas in the velocity interval $dv$ is constant $W(v) dv=C_W dv_m$, where $C_W$ is a constant.
Changing variable from the velocity along the tube to that along the line of sight,
this weighting function (up to a constant) is, by using $v_D=v_{\rm m} \cos\psi$,
\begin{equation}
W(v_{\rm m})dv = W(v_D) dv_D =
\frac{C_W} {\cos \psi} {dv_D} , \qquad (v_0 \cos \psi) < v_D < (v_{\rm max} \cos \psi).
\label{eq:W}
\end{equation}
At each orbital angle $\theta$ we sum the contribution to absorption at
Doppler shift $v_D$ over all tubes,
\begin{equation}
A(v_D)=\sum_{\rm tubes} W(v_D)  \sin \psi .
\label{eq:sum}
\end{equation}
The intensity is taken to be
\begin{equation}
I(v_D,\theta)=1-K_a A(v_D,\theta) ,
\label{eq:I}
\end{equation}
where $K_a$ is a constant of the toy-model that takes care of units and the condition $0<I<1$.
We set here$K_a=1/ \rm{max}(A)$, because we do not compare to the absorption equivalent width.

Our model is actually a toy-model. It assumes a simple geometry of the
absorbing  material, i.e., a rotated conical shell with a varying opening
angle, and a mass distribution within each tube that is constant with the velocity.
The model contains two types of parameters:
(1) Those that are given in the literature.
Such are the binary parameters and the conical
shape with its opening angle $\phi_a$.
These parameters are more or less in consensus.
The orientation of the semimajor axis (the periastron longitude $\omega$) is
controversial, and we take the value from our previous papers $\omega=90^\circ$
(secondary toward us at periastron).
(2) Parameters that are unique to our toy-model.
Such is the value of $v_{\rm max}$
which is constraint to be $v_{\rm max} \la v_{\rm 2,\infty}=3000 \km \s^{-1}$.
For these parameters we find that a good general fit can be
obtained for $v_{\rm max}=2300 \km \s^{-1}$.
The value of the absorption coefficient $K_a$ has a small influence on our
conclusions, and it serves only to give the general form of the absorption profile.

The assumption that the conical shell reaches its asymptotic
opening angle at large distances breaks down as the system approaches periastron.
The reason is that the relative velocity of the two stars is
no longer much smaller than the primary's wind speed.
This causes the winding-up of the conical shell into a spiral structure in the
equatorial plane, e.g., as shown in the numerical study by Okazaki et al. (2008b).
Only a close region near the binary system reaches this limiting angle.
Namely, a smaller region, but much denser, will contribute to the absorption at the blue edge.
However, contribution to the continuum near $1 \mum$ comes from the stellar wind
and hot dust from closer regions to the binary system (Kashi \& Soker 2008a).
Therefore, the much smaller region of the conical
shell can still absorb a detectable fraction of the continuum.
Our toy-model does not allow us to make any quantitative prediction.
For that a 3D numerical code is required. Nevertheless, our toy-model does allow us to
reach the two goals, as mentioned in the last paragraph of section \ref{sec:intro}.

\section{THE BLUE ABSORPTION WING}
\label{sec:results}

\begin{figure}[!ht]
\resizebox{0.89\textwidth}{!}{\includegraphics{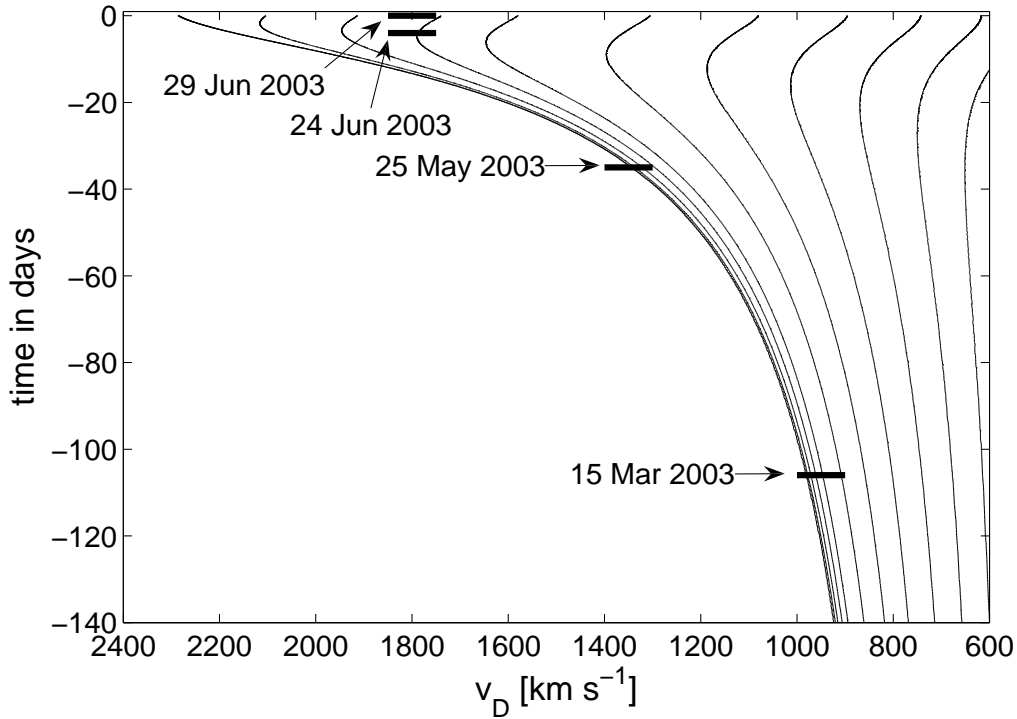}}
\caption{\footnotesize
A contour map of the intensity $I$ as given in equation \ref{eq:I}, in the Doppler shift$-$time plane.
The blue Doppler shift is given in unit of $\km \s^{-1}$, and the vertical axis indicates days before periastron (phase 0).
The levels of the contours are $I=1, 0.95, 0.9, 0.85, 0.8, 0.7, 0.6$, and so on, from left to right.
The edges of the blue absorption wing at four epochs from Damineli et al. (2008b) are marked.
The length of each of the drawn lines corresponds to $100 \km \s^{-1}$, our estimate of the uncertainty in determining the edge of the absorption.
We predict the edge Doppler shift to reach a maxima 0-20 days (according to our parameters) before periastron,
when a maximum in $v_D$ appears according to our model.
}
\label{fig:abs}
\end{figure}

In Figure \ref{fig:abs} we present our results by a contour map of the intensity $I$,
as given in equation \ref{eq:I}, in the Doppler shift$-$time plane.
The blue Doppler shift $v_D$ is given in unit of $\km \s^{-1}$,
and the vertical axis indicates days relative to (before) periastron (phase 0 at $t=0$).
The levels of the contours are $I=1, 0.95, 0.9, 0.85, 0.8, 0.7, 0.6$, and so on,
from left to right, where $I=1$ indicates the edge of the absorption profile.
Despite the constant weighting function $W$ that we used,
in reality we expect that the amount of helium blown with high velocities corresponding to
$I \simeq 0.9-1$ will be too small to be detected.
We have also estimated the velocity edges of the absorption wing at four epochs
from Damineli et al. (2008b) observations; these are indicated by four horizontal error-bar
lines,.
The somewhat noisy data made it difficult to pinpoint the exact edge of the wing,
and therefore we could only estimate it to an accuracy of $\sim 100 \km \s^{-1}$.
Each line is centered at the approximated edges of the absorption wing,
and extends to $\pm 50 \km \s^{-1}$ to each direction.
At early times there is no noticeable differences between the contour lines in the range $I=0.85-1$,
and all nicely fit the two observation equally well.
Very close to periastron, the line $I=0.85$ fits the observations better.
However, during this time we expect the collapse of the conical shell, such that
no fresh gas will be accelerated to high velocities.
We cannot make an accurate prediction so close to periastron passage.
Over all, we consider the lines $I=0.85-1$ to be a very good fit to observations.

We modified the parameters in our model to fit
the observations of Damineli et al. (2008b) from the 2003 event.
With this fitting we learn about two properties of the models.
(1) We find that the orientation we use, where the secondary is toward us at periastron
($\omega =90^\circ$) can account for the development of a wide blue absorption wing
starting several months before periastron passage, if the absorbing material is
within this conical shell.
(2) According to our model, the maximum observed blue shift of the wing
is reached  $\sim 5 \days$ before periastron passage.
The observed maximum blue shift might occur at a somewhat different time for three reasons.
Firstly, the binary and winds parameters might be somewhat different than those we used.
Secondly, we did not take into account the winding of the conical shell; we only
considered the approximate direction of its axis by calculating the angle $\delta \phi$.
Thirdly, close to periastron the conical shell is likely to collapse onto the secondary
(Soker 2005; Kashi \& Soker 2009).

With the only four available observations for the fitting,
with the present possible degree of accuracy, and according to our parameters,
we expect the maximum blue shifted absorption to occur $0-20 \days$ before periastron passage,
i.e., late December or early January. There are two main reasons for that we cannot be more accurate.
Firstly, we cannot treat properly the conical shell as the system approach periastron.
Secondly, we expect the colliding wind region to collapse onto the secondary near periastron; we
did not consider this process here either.

\begin{figure}[!ht]
\resizebox{0.75\textwidth}{!}{\includegraphics{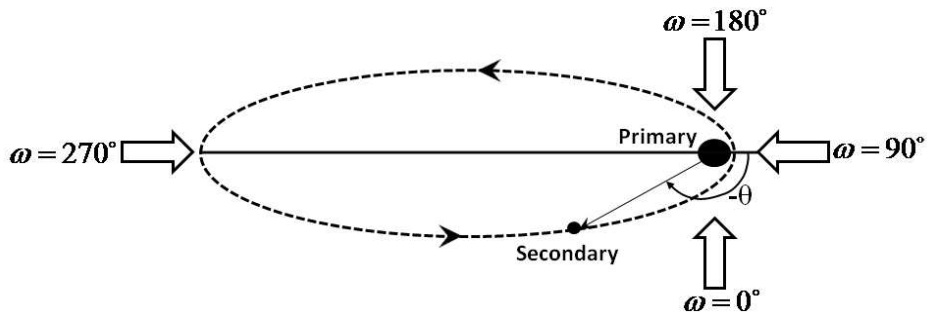}}
\caption{\footnotesize
The fours semimajor orientations studied in the paper. In all cases the
inclination angle (the angle between line of sight and a line perpendicular to the orbital plane)
is $i = 41^\circ$. Our favorite model has $\omega=90^\circ$, for which
the blue absorption wing is drawn in figure \ref{fig:abs}. }
\label{fig:omega}
\end{figure}

Our fundamental assumption is that the absorber of the blue wing of the He~I~$\lambda10830 \rm{\AA}$
line resides in the conical shell formed by the colliding winds.
This assumption works quite well with the semimajor orientation $\omega=90^\circ$,
where the secondary is closest to the observer at periastron passage.
We now show that other orientations that are popular in the literature cannot account,
not even qualitatively, for the behavior of the blue wing, if our fundamental assumption holds.
We followed the calculations presented in Figure \ref{fig:abs}, but for
three different semimajor axis orientations, as drawn in Figure \ref{fig:omega}:
$\omega=0$: The semimajor axis is perpendicular to the line of sight and the secondary
is closer to the observer before the event.
$\omega=180^\circ$: The semimajor axis is perpendicular to the line of sight and the secondary
is closer to the observer after the event.
$\omega=270^\circ$: The secondary is closest to the observer at apastron, opposite to our favorite
orientation of $\omega=90^\circ$ (Figure \ref{fig:abs}).

\begin{figure}[!ht]
\resizebox{0.69\textwidth}{!}{\includegraphics{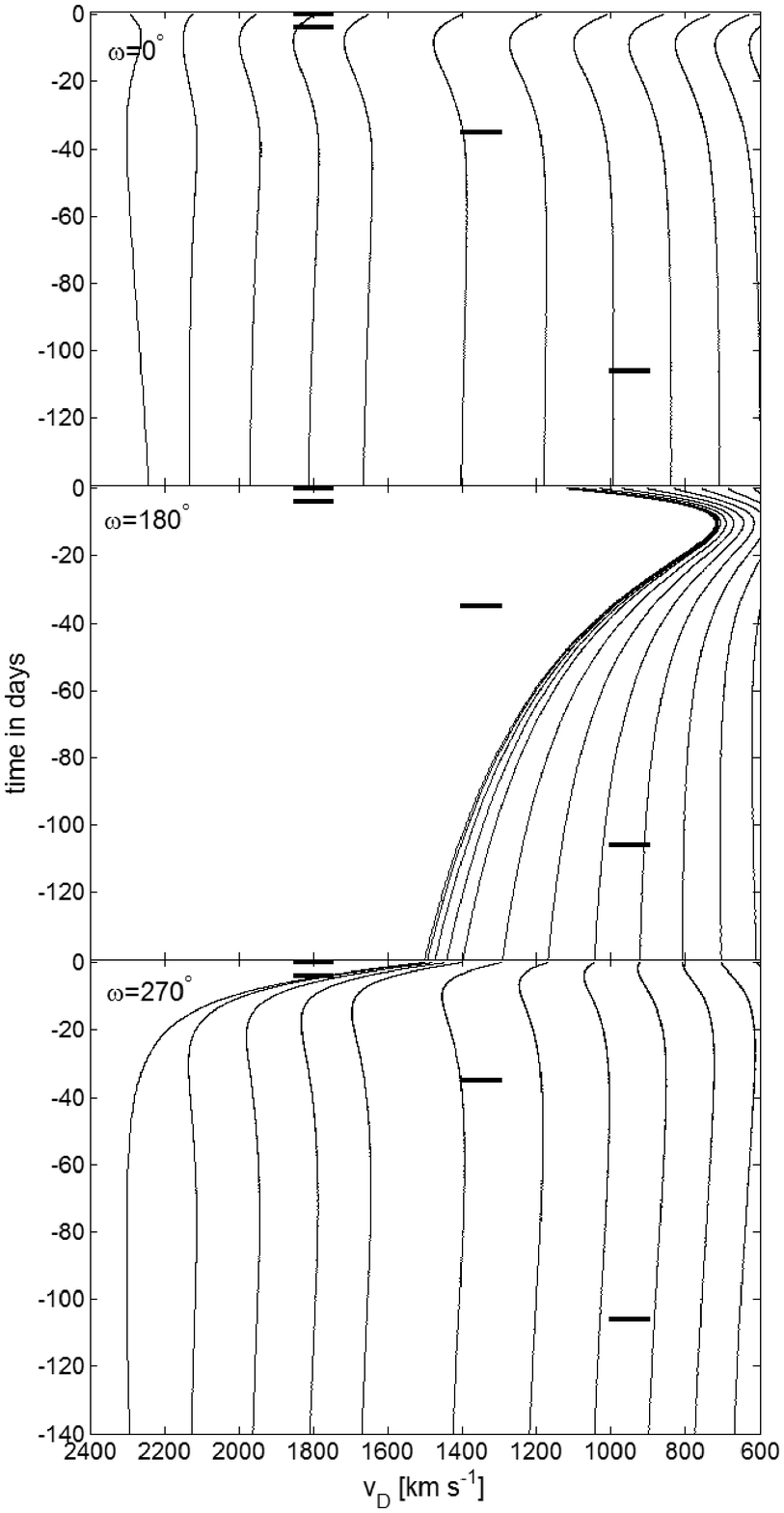}}
\caption{\footnotesize
Same as in Figure \ref{fig:abs}, but for different values of $\omega$,
as schematically depicted in Figure \ref{fig:omega} .
None of the other orientations can account for the observed absorption wing
under our fundamental assumption that the absorber reside in the conical
region formed by the collision of the two winds.
In particular, the opposite orientation, $\omega=270^\circ$, yields an
entirely opposite behavior to observations, with minimum absorption at periastron.
}
\label{fig:otheromega}
\end{figure}

The results for the expected Doppler shifts of the absorption wing are presented
in Figure \ref{fig:otheromega}, together with the fours observations from Damineli et al. (2008b).
It is clear that none of the other orientations can account for the observed absorption
wing, not even qualitatively and just for the two early observations.

\section{SUMMARY AND PREDICTION}
\label{sec:es}

We study the blue absorption wing of the He~I~$\lambda10830 \rm{\AA}$ P Cygni profile of $\eta$ Car.
The two winds of the two stars collide, and the post shocked gas of the two winds flow
on both sides of a surface (the discontinuity surface) that has a
pseudo-conical structure: the conical shell.
The shocked secondary's wind accelerates part of the shocked primary's wind to high velocity
(Figure \ref{fig:colliding}).
This gas, and the segments of the shocked secondary's wind which cool to a low temperature near
the contact discontinuity (Figure \ref{fig:colliding}), are assumed to be
responsible for the blue absorption wing.
This is the fundamental assumption of our model.
We use our previous results and assume an orientation where the secondary is toward
the observer at periastron ($\omega=90^\circ$; see Figure \ref{fig:omega}).
For the conical shell we built a toy model (Figure \ref{fig:cone}).

The absorption profile, up to a scaling factor, is calculated according to equations
(\ref{eq:sum}) and (\ref{eq:I}), and the results are presented in Figure \ref{fig:abs}.
We are interested in the bluest part of the absorption profile, where intensity is lower;
in our scaled units these are the contours in the range $I \simeq 0.8-1$.
Using our fundamental assumption, and a toy model for the conical shell of the colliding
winds, we showed that with our orbital orientation we can account for the
appearance of the wide blue wing several months before periastron.
Other semimajor orientations $\omega$, cannot reproduce the results under our
fundamental assumption (figure \ref{fig:otheromega})
This is our main result, namely, that if the absorber responsible for the
blue wing of the He~I~$\lambda10830 \rm{\AA}$ line reside in the winds
collision region, then only the $\omega \simeq 90^\circ$ can account for the
blue wing of this line.

Our results also predict that the Doppler shift $v_D$ (the edge of the profile)
will reach a maximum $0-3~\rm{weeks}$ before periastron passage.
Since close to periastron the conical shell starts to collapse onto the
primary, and even before it experiences the effect of wrapping,
it is hard to pinpoint the exact time of this maximum.
Nevertheless this maximum should be observed before the event.

\acknowledgements
We thank Augusto Damineli for providing us the observational data
for the He~I~$\lambda10830 \rm{\AA}$ line, and an anonymous referee for helpful comments.
This research was supported by the Asher Space Research Institute in the Technion.


\end{document}